\documentclass[aps,pra,twocolumn,showpacs,superscriptaddress,groupedaddress,reprint]{revtex4}
\usepackage{amsthm}
\usepackage{newlfont}
\usepackage{graphicx}
\usepackage{epstopdf}
\usepackage{appendix}
\usepackage{hyperref}
\usepackage{textcomp}
\usepackage{appendix}
\usepackage{multirow}	
\begin{document}
\title{\textbf{Disorder Induced Vortex Lattice Melting in Bose-Einstein Condensate}}
\author{T. Mithun$^1$, K. Porsezian$^1$ and Bishwajyoti Dey$^2$\\
\small \textsl{$^1$Department of Physics, Pondicherry University, Puducherry 605014, India.\\
$^2$Department of Physics, SP Pune University, Pune 411007, India.}}
\begin{abstract}
 We study the vortex lattice dynamics in presence of single impurity as well as random impurities or disorder. We show that in presence of a single impurity the vortex lattice gets distorted and the distortion  depends on the position of the single impurity with respect to the positions of the vortices in the impurity free Abrikosov vortex lattice and also the strength of the impurity potential. We then show that a new type of giant hole with hidden vortices inside it can be created in the vortex lattice by a cluster of impurities. In presence of random impurity potential or disorder the vortex lattice melts. We show that the vortex lattice also melts in presence of pseudorandom potential generated by the superposition of two optical lattices. The  absence of long-range order in the melted vortex lattice is demonstrated from the structure factor profile and the histogram of the distances between each pair of the vortices. 
\end{abstract}
\maketitle
\pagestyle{myheadings}
%\section{Introduction}
The study of Bose-Einstein Condensates (BEC) in random potential has got much attention because in this system it is possible to display in a controlled way the interplay between interaction and disorder. In real-life systems random impurities are unavoidable and contrary to expectation, strong effects survive averaging over the disorder. Recently it has been shown that long-range correlations can be enhanced when disorder breaks the continuous symmetry of a system.  Such a random-field-induced-order can be realized in ultracold atoms in optical lattice \cite{Guillamon1}. Random potential in BEC can be created by optical speckles \cite{Fort, Lye, Clement}. Bose-Einstein condensates in random potential has emerged as an ideal ground for studying several problems such as Anderson localization \cite{Billy}, superfluid behavior \cite{Konenberg}, superfluid-Mott-insulator transition \cite{Chen, Greiner}, Bose glass and their microscopic properties \cite{Giamarchi, Dalfovo}, superconductivity and quantum magnetism etc \cite{Palencia}. 
 
A rapidly rotating BEC creates highly ordered quantized triangular vortex lattice \cite{Engels} which  mimics the Abrikosov vortex lattice in a type II superconductors placed in magnetic fields \cite{Abrikosov}. For a review on  the experimental and theoretical studies of vortex lattices in rotating BEC, see  \cite{Fetter}. When the rotating BEC is placed in a co-rotating optical lattice the vortex lattice gets pinned to the optical lattice \cite{Sato}. Recently, we have shown that the hidden vortices can also be pinned by the optical lattice \cite{dey14}.  The study of melting of vortex lattice in BEC is an important problem as it provide an ideal system for understanding the mechanism of lattice melting of the two-dimensional systems in general. In presence of significant thermal fluctuation, in particular for high-temperature superconductors, the vortex lattice melts into vortex liquid and in presence of disorder the vortex lattice undergoes a transition to a vortex glass state \cite{Rosenstein}. This transition has been investigated in details for type II superconductors \cite{Rosenstein,Safar,Koshelev}. In an interesting recent paper a direct observation of melting in a two-dimensional superconducting vortex  lattice using scanning tunneling spectroscopy is reported \cite{Guillamon2}.  Even though the superfluids (BEC and liquid helium) and superconductors  are very different systems, the vortex dynamics in these systems show similar behaviour. For example, the equilibrium vortex lattice configuration in both the rotating BEC and type-II superconductors in applied magnetic field is Abrikosov triangular lattice. Similarly, the dynamics of vortices in presence of impurity (pinning of vortices) in both the systems also show similar behaviour. The similarity in the vortex dynamics between the superfluid and the type-II superconductors is due to the fact that the transverse force or the Magnus force acting on a quantized vortex in a superfluid and type-II superconductor has the same form ${\textbf f} = \rho {\textbf K}\times {\textbf V}$ where $\rho$ is the mass density, ${\textbf K}$ is the quantized circulation vector and ${\textbf V}$ is the vortex velocity \cite{Thouless, Geller}. Studying vortex dynamics in superconductor is difficult because of the presence of natural impurities in the system. Therefore BEC provides a system where it is possible to display in a controlled way the interplay between interaction and disorder on the vortex dynamics. In reference \cite{Gifford} the dislocation-mediated thermal melting of the two-dimensional superfluid vortex lattice is well explored theoretically, while in \cite{Cooper,Sinova,Regnault,Snoek1} melting due to quantum fluctuations are reported.  It has been shown that the melting of vortex lattice in BEC can take place due to  fast rotation of the condensate  \cite{Kasamatsu1}. The anisotropic compression have also been used to melt vortex lattice in BEC \cite{Engels1}. The investigation of vortex lattice melting in BEC trapped in quartic radial potential shows that vortex liquid can form in the center of BEC instead of at the outer edge of the pancake BECs \cite{Snoek2}.  The quantum melting of the vortex lattice in a rapidly rotating quasi-two-dimensional BEC due to the cooperative ring exchange mechanism is studied in \cite{Ghosh}. The understanding of melting of the two-dimensional systems have impact across several research fields \cite{Guillamon2} and vortex lattices are considered as ideal systems where mechanism of such processes can be investigated.

In this article we present a new mechanism of  vortex lattice melting in BEC which is induced by the presence of random impurities or disorder in the system. For this we study numerically the effects of random potential on the vortex lattice of a rotating BEC.   Disorder is introduced in the system by the external potential associated with the random impurities. In order to understand the role played by the impurities we also study the effect of a single impurity on the vortex lattice in BEC.  Our numerical calculations based on the Gross-Pitaevskii equation (GPE)  show that a single vortex can be trapped by a single impurity. The vortex lattice gets distorted in presence of a single impurity. The extent of the vortex lattice distortion depends on the position of the impurity (commensurate or incommensurate with respect to the undistorted Abrikosov vortex lattice) as well as the strength of the impurity potential. We also show that a cluster of impurities can create a giant hole containing hidden vortices in the vortex lattice. Finally, we show the disorder induced vortex lattice melting in presence of random impurity potential and also the pseudorandom potential generated by the superposition of two optical lattices. 
  %----------------------------
    \begin{figure*}[pht] 
       \vspace{0pt}
  \includegraphics[width=\textwidth,height=7.0cm]{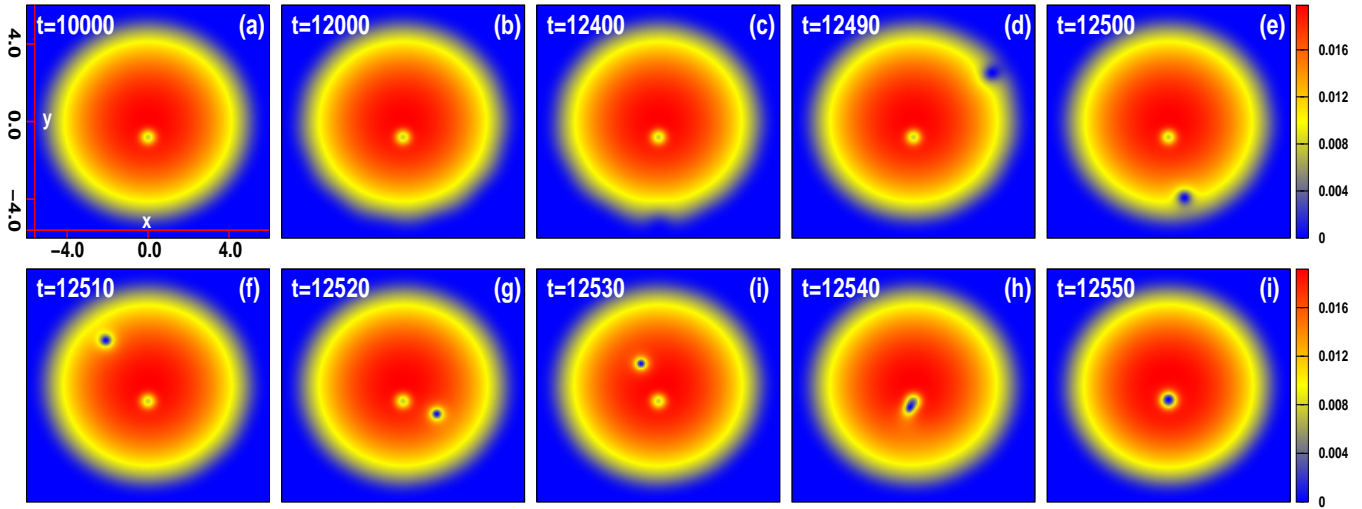}
     \vspace{0pt}
 \caption{\label{Fig. 1}{\footnotesize (Color online) (a)-(i) Condensate density $|\psi|^2$ showing the time evolution of the single vortex position. The vortex gets pinned to the single impurity placed at the position $(0, -0.8)$ following a spiral path. Here $V_0=20$ and $\Omega=0.479$.}}
%      \vspace{-15pt}
%\label{1}
  \end{figure*}
    %----------------------------
      \begin{figure*}[pht] 
      \centering
       \vspace{0pt}
  \includegraphics[width=\textwidth,height=7.0cm]{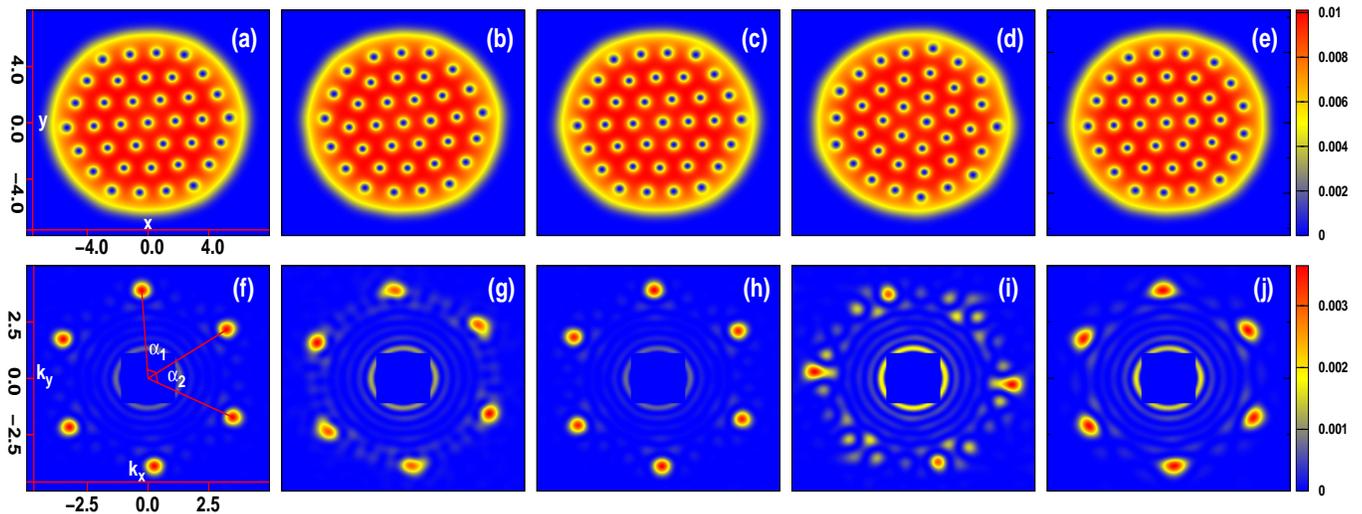}
     \vspace{0pt}
 \caption{\label{Fig. 2}{\footnotesize (Color online) Condensate density $|\psi|^2$ depicted in upper panel shows the vortex lattice, while the lower panel shows the structure factor profile of each vortex lattice situating above its top. Here (a) and (f) are $|\psi|^2$ and its structure factor without impurity, while (b)-(e) show $|\psi|^2$ for single impurity which is kept at the positions $(x_0=1.8, y_0=0.12)$,$(x_0=0.4, y_0=0.023)$, $(x_0=0.9, y_0=0.06)$, and  $(x_0=1.5, y_0=0.1)$ respectively, and (g)-(j) are the corresponding structure factor profile. Angles $\alpha_1$ and $ \alpha_2$ are measured as marked in (f).  Here $V_0=5$ and $\Omega=1.1$. }}
%      \vspace{-15pt}
%\label{1}
  \end{figure*}
        %----------------------------
    \begin{figure}[pht] 
       \vspace{0pt}
  \includegraphics[width=0.5\textwidth,height=6.0cm]{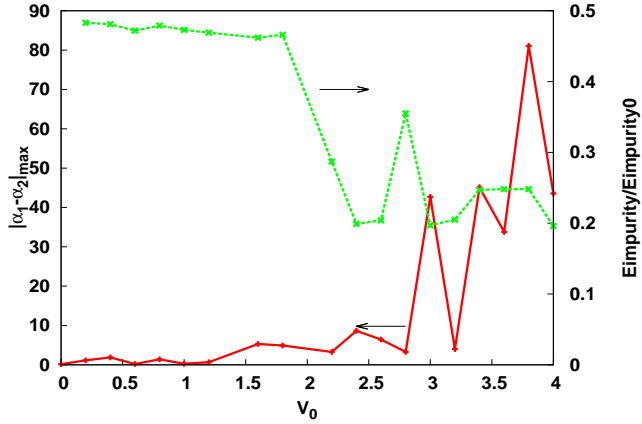}
     \vspace{0pt}
 \caption{\label{Fig. 3}{\footnotesize (Color online) Variation of the angular distortion $|\alpha_1-\alpha_2|_{\textrm {max}}$ (red line) and the lattice energy (green dots) with $V_0$, where $V_0$ is the strength of the single defect for $\Omega=1.1$. Single defect is at the position (0.9, 0). Here $E_{\textrm {impurity0}}$ is the impurity potential energy when $\Omega=0$.}}
%      \vspace{-15pt}
%\label{1}
  \end{figure}
  %----------------------------
        \begin{figure}[phb] 
       \vspace{0pt}
  \includegraphics[width=7.0cm,height=6.0cm]{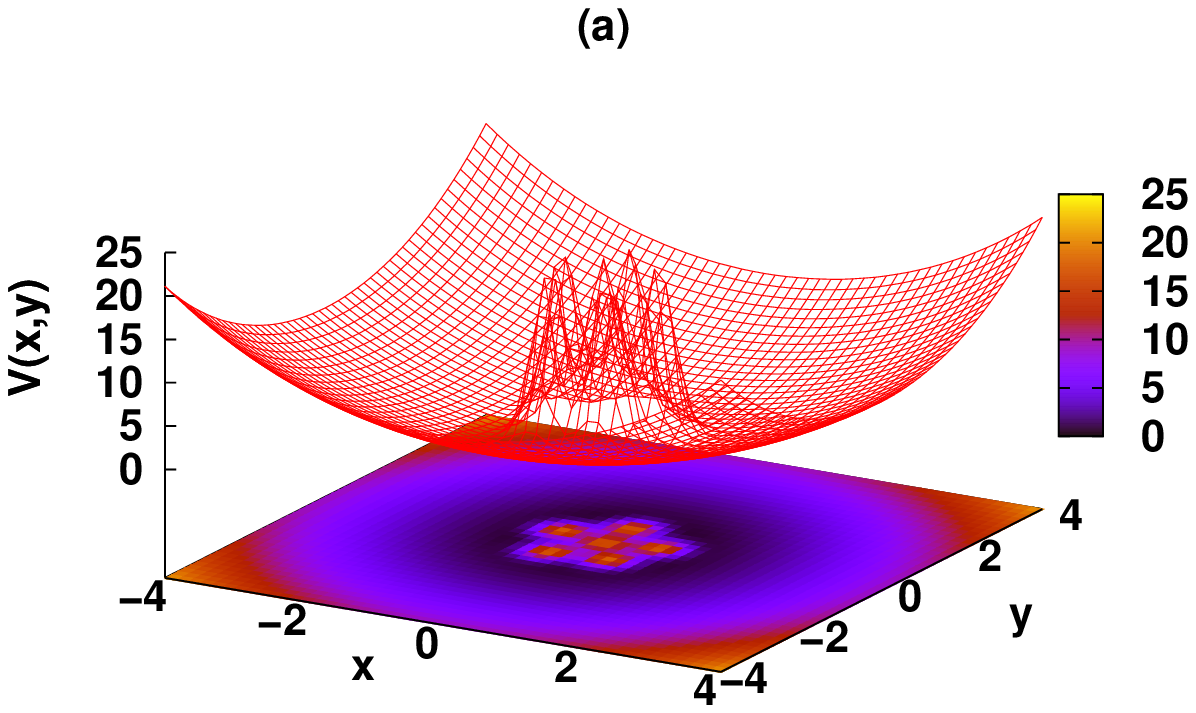}\newline
    \includegraphics[width=4.0cm,height=3.5cm]{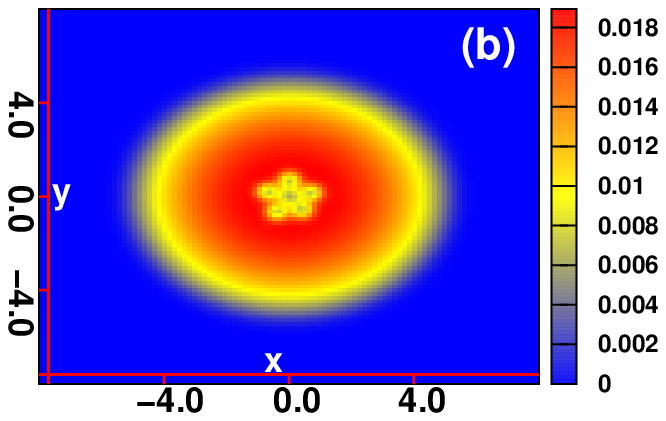}
     \vspace{0pt}
 \caption{\label{Fig. 4}{\footnotesize (Color online) (a) The potential of a cluster of 6 impurities placed in the harmonic plus quartic trap potential, and (b) the condensate density $|\psi|^2$ in absence of rotation. Here $V_0=20$.}}
%      \vspace{-15pt}
%\label{1}
  \end{figure}
  %----------------------------
        \begin{figure}[phb] 
       \vspace{0pt}
  \includegraphics[width=8.5cm,height=7.5cm]{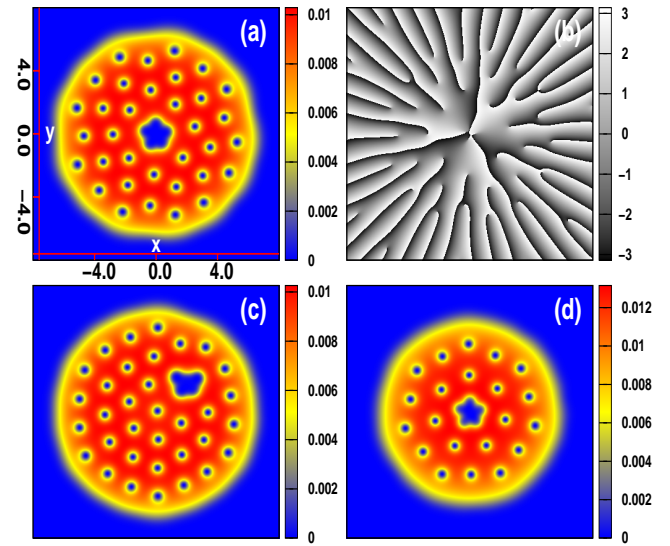}
     \vspace{0pt}
 \caption{\label{Fig. 5}{\footnotesize (Color online) (a) Giant hole in the condensate density $|\psi|^2$ for a cluster of impurities  kept at the center for $\Omega=1.1$, (b) corresponding phase profile, (c) giant hole for a cluster of impurities placed away from the center for $\Omega=1.1$, and (d) giant hole for cluster of impurities kept at the center for $\Omega=0.95$. Here $V_0=20$.}}
%      \vspace{-15pt}
%\label{1}
  \end{figure}
      %----------------------------
            \begin{figure}[pht] 
       \vspace{0pt}
  \includegraphics[width=8.5cm,height=7.5cm]{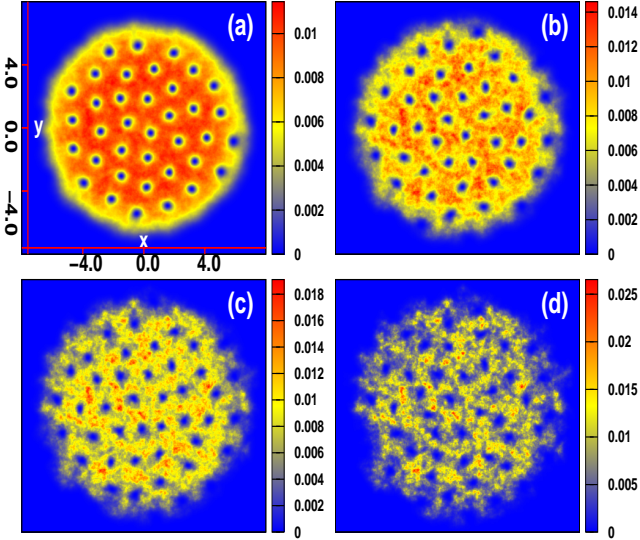}
     \vspace{0pt}
 \caption{\label{Fig. 6}{\footnotesize (Color online) Condensate density $|\psi|^2$ for various strength $V_0$ of the random impurity potential. (a) $V_0 = 10$, (b) $V_0 = 30$ , (c) $V_0 = 50$, and  (d)  $V_0 = 80$. Here $\Omega = 1.1$.}}
%      \vspace{-15pt}
%\label{1}
  \end{figure}
      %----------------------------

        \begin{figure}[pht] 
       \vspace{0pt}
  \includegraphics[width=5.0cm,height=4.2cm]{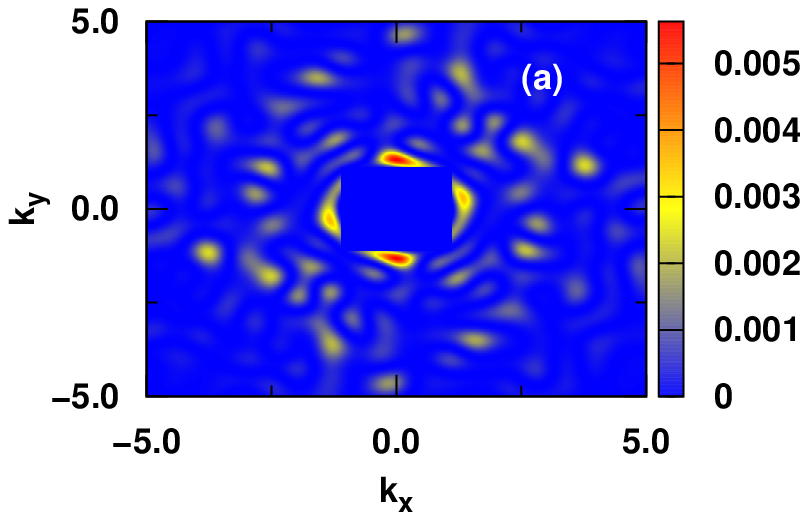}\newline
    \includegraphics[width=8.0cm,height=6.5cm]{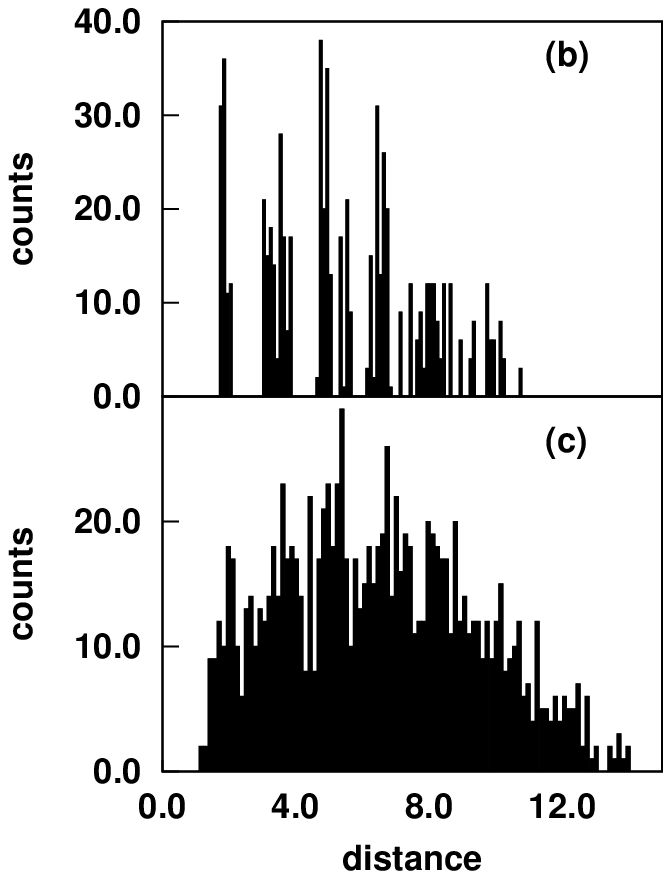}
     \vspace{-2.0pt}
 \caption{\label{Fig. 7}{\footnotesize (Color online) (a) The structure factor profile showing the degree of distortion of the melted vortex lattice of Fig. \ref{Fig. 6}(d), (b) the histogram plot for Fig. \ref{Fig. 2}(a), and (c) histogram plot for Fig. \ref{Fig. 6} (d).}}
%      \vspace{-15pt}
%\label{1}
  \end{figure}
        %----------------------------
\begin{figure}[pht] 
       \vspace{0pt}
 \includegraphics[width=8.0cm,height=3.5cm]{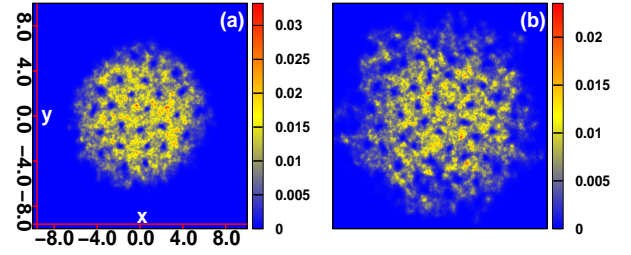}
     \vspace{-2.0pt}
 \caption{\label{Fig. 8}{\footnotesize (Color online) Condensate density $|\psi|^2$ showing melted vortex lattice (a) for the anharmonic trap potential, and  (b) for the harmonic trap potential. Here $\Omega = 0.9$ and $V_0 = 80$}}
\end{figure}
%---------------------------------------
        \begin{figure}[pht] 
       \vspace{0pt}
  \includegraphics[width=8.0cm,height=3.5cm]{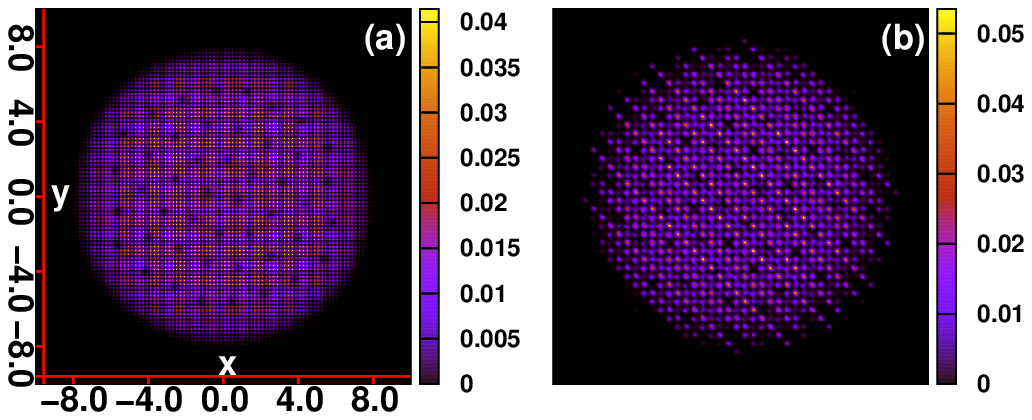} \newline
    \includegraphics[width=8.0cm,height=6.5cm]{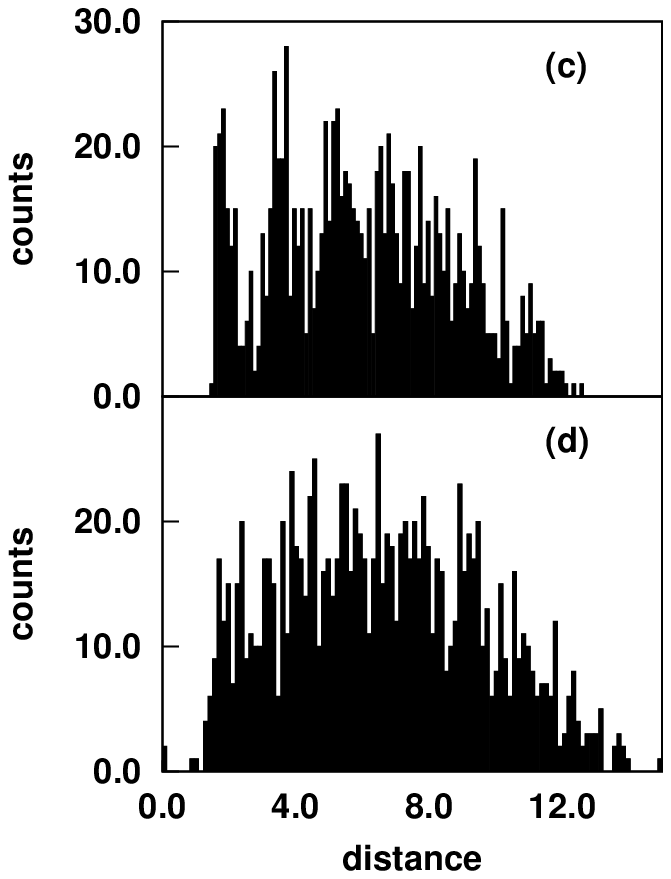}
     \vspace{-2.0pt}
 \caption{\label{Fig. 9}{\footnotesize (Color online) Condensate density $|\psi|^2$ showing melted vortex lattice (a) for biochromatic potential,  and (b) for combined potential (main lattice+secondary lattice), (c) histogram plot for Fig. (a), and (d) histogram plot for Fig. (b).}}
%      \vspace{-15pt}
%\label{1}
  \end{figure}
    %----------------------------
%\section{Theoretical Model}
The dynamics of the system is described by the two-dimensional (2D) dimensionless  time-dependent Gross-Pitaevskii Equation (2DGPE) 
$$\small  (i-\gamma)\psi_{t}=[-\frac{1}{2}(\partial_{x}^2+\partial_{y}^2 )+V(x,y)-\mu+p|\psi|^2-\Omega L_{z}]\psi \eqno{(1)}$$

\noindent 
where $V(x,y)=V_{trap}+V_{imp}$, $\Omega$ is the rotational frequency and $L_{z}$ is the angular momentum in z direction. We consider the condensate is trapped in a harmonic plus quartic trap potential $V_{trap}=\frac{1}{2}(r^2 + \lambda r^4)$. The advantage of harmonic plus quartic trap potential over the harmonic potential is that in the former case the condensate can be rotated with rotational frequency larger than the trap frequency ($\Omega >1$) \cite{Fetter1, Kasamatsu2}. However we show that the impurity-mediated vortex lattice melting is possible even for $\Omega < 1$.  Further we have shown that the vortex lattice melting is also possible for harmonic trap potential. $V_{imp}$ is the impurity potential. For a single nonmagnetic impurity at position $(x_0,y_0)$ we take the impurity potential as $V_{imp}=V_0 \exp[-\big((x-x_0)^2+(y-y_0)^2\big)/(\sigma/2)^2]$ where $V_0$ is the strength of the impurity potential and $\sigma$ is width of the potential. Such a single defect can be created experimentally \cite{Fort}. For random nonmagnetic impurities the impurity potential or disorder is defined by an independent random variable uniformly distributed over $[-V_0,V_0]$ at each spatial position $\textbf {r}$, where $V_0$ denotes the strength of the disorder \cite{Ghosal}.  Experimentally it is  possible to create  random impurities using laser speckle method \cite{Palencia}. Another way to create such disorder or pseudorandom potential is by superposing two optical lattices of incommensurate periods. The disorder in this case is represented by the bichromatic potential $V_{imp} = V_0[\cos(2\pi\beta x + \phi) + \cos(2\pi\beta y + \phi)] $, where $\beta = k_2/k_1$ is the irrational ratio of the wave vectors of the two optical lattices and $\phi$ is their phase differences. Although the potential is deterministic, it mimics disorder in finite-size system \cite{Palencia}. Pseudoandom potential can also be created by adding a second additional lattice to the main optical lattice $V_{imp} = V_0[\cos^2(kx) + \cos^2(ky)] + V_1[ \cos^2(\textbf{k}_1 . \textbf{r})+\cos^2(\textbf{k}_2 . \textbf{r})]$ \cite{damski}. The randomness of the potential is determined by the ratio of the wavelengths of the main and additional optical lattice. The origin of the dissipation term ($\gamma$) in Eq. (1) is due to the presence of thermal component present in the trap or in other words it is related to the collision of the noncondensed atoms with the condensate atoms. To include dissipation, the dynamical equation given by the Gross-Pitaevskii equation (GPE) is modified by including the $\gamma$-term in the equation. For the derivation of the modified GPE see \cite{choi}. The value of the parameter $\gamma$ has been obtained by fitting theoretical results with experiments to be $\gamma = 0.03$ \cite{choi}. Thereafter this particular value of the parameter $(\gamma =0.03)$ have been used by several authors \cite{dey14, Kasamatsu1, Kato, Kasamatsu3, Tsubota} etc. for studying vortex dynamics in BEC.  The damping term helps faster convergence to the equilibrium state. The small variation of $\gamma$ value only effects the convergence time. In our system this term is introduced for faster convergence of the system to the equilibrium vortex lattice state. In Eq. (1), the spatial coordinates, time, condensate wave function, rotational frequency and energies are in units of $a_{0}$, $\omega_{\perp}^{-1}$, $a_{0}^{-3/2}$, $\omega_{\perp}$ and $\hbar\omega_{\perp}$ respectively, where $a_{0}=\sqrt{\hbar/m \omega_{\perp}}$.

The 2DGPE (Eq. (1)) is solved numerically using the split-step Crank-Nicolson method. In our simulation, we set small spatial step $\Delta x$ = $\Delta y$ = 0.08 and time step $\Delta t$ = 0.001 for Fig. \ref{Fig. 1}, while $\Delta x$ = $\Delta y$ = 0.04 and $\Delta t$ = 0.0005 for Figs. \ref{Fig. 2}-\ref{Fig. 8}. The parameters are chosen from the experiments on $^{87}Rb$ \cite{Kato, Williams}. The dissipation parameter is set to $\gamma =0.03$ and $\lambda=0.01$. Eq. (1) is propagated in imaginary time by replacing $t$ as $t \rightarrow - it$ until the ground state solution is achieved and thereafter it is propagated in real-time. In order to conserve the norm of the wave function for nonzero value of $\gamma$, the chemical potential is treated as time-dependent and is adjusted at each time step.  

 We characterize the distortion of the vortex lattice in presence of impurity in terms of (i) the structure factor $S(\textbf{k})=\int d\textbf{r} |\psi(\textbf{r},t)|^2 e^{i \textbf{k}.\textbf{r}}$ (ii) the impurity potential energy $\langle V_{imp}\rangle $ \cite{Sato, Guillamon2} and (iii) the histograms of distances between each pair of vortices  \cite{Engels1}. We calculate the variation of these quantities with the strength  of the impurity potential. 

To show the pinning of the single vortex by an impurity we study the dynamical evolution of the rotating condensate  in presence of an impurity or pinning center. We track the positions of the single vortex at different times of the dynamical evolution. Fig. \ref{Fig. 1}  shows the positions of the single vortex at different times in presence of a single impurity at $(0, -0.8)$. From the figures in the two panels we can see that the single vortex enters the condensate at the top  and finally gets pinned to the impurity by following a spiral path. The pinning potential prevents the vortex movement to other places by providing the increased effective potential as discussed in \cite{Isoshima}. Similar pinning of a single vortex (flux line) by a pinning center also takes place in type-II superconducting films \cite{Ryuzo}. We now show how the vortex lattice gets distorted as the position of the single impurity is varied. In absence of impurity the vortex lattice in rotating BEC is a triangular Abrikosov lattice where the vortex lattice positions have translational periodicity or long-range order as shown in Fig. \ref{Fig. 2}(a). In addition to condensate density profile we have also plotted the density structure factor profile. The structure factor provide us information about the periodicity of the condensate density. For Abrikosov triangular lattice, the structure factor profile shows periodic peaks of regular hexagonal lattice \cite{Sato}. The structure factor profile for Fig. \ref{Fig. 2}(a) is shown in Fig. \ref{Fig. 2}(f) where we can see periodic peaks of regular hexagonal structure.  Figs. \ref{Fig. 2}(b-e) show the vortex lattice pattern when a single impurity  position is varied from a commensurate position (Fig. \ref{Fig. 2}(b)) to various incommensurate positions (Figs. \ref{Fig. 2}(c-e)).  Here commensurate position of the single impurity means it coincides with a vortex position of the undistorted Abrikosov lattice Fig. \ref{Fig. 2}(a). In presence of impurity the vortex lattice gets disordered  and the long-range order or lattice translational invariance is lost. This is reflected in the corresponding structure factor profiles shown in Figs. \ref{Fig. 2}(g-j) respectively. From these figures we can see that the distortion of the vortex lattice depends on the position of the impurity and the distortion is minimum when the impurity position commensurate with the undistorted vortex lattice positions. The vortex lattice distortion is maximum when the impurity is placed in the center of the two vortices of the Abrikosov vortex lattice (Fig. \ref{Fig. 2}(d)). From the corresponding structure factor profile (Fig. \ref{Fig. 2}(i)) we can see that there are no periodic peaks and no regular hexagonal lattice structure. Fig. \ref{Fig. 3} shows the variation of the angular difference $|\alpha_1-\alpha_2|_{\textrm {max}}$ and the impurity potential energy E$_{\textrm {impurity}}$ = $\langle V_{imp}\rangle $ of the distorted vortex lattice with the strength of the single impurity potential. The angles are defined in Fig. \ref{Fig. 2}(f). Here $|\alpha_1-\alpha_2|_{\textrm{max}}$ represents the maximum of the difference in angles. The increase in the angular difference  $|\alpha_1-\alpha_2|_{\textrm {max}}$ show that the vortex lattice gets increasingly distorted  with the strength of the impurity potential $V_0$ (red line). Similarly the impurity potential energy E$_{\textrm {impurity}}$ initially decreases quite rapidly with increasing $V_0$ (green dots) and finally it attains nearly a constant value showing that the vortex lattice has got pinned to the impurity \cite{dey14}. 

It is known that a `giant hole' created in BEC can be used to study controlled circular superflow and phase slippage in superfluids and also the transverse Tkachenko modes in the vortex lattice \cite{Coddington}. Kasamatsu {\it et al} created a giant hole in the condensate density of a BEC trapped in a harmonic plus quadratic potential \cite{Kasamatsu2} by rotating the condensate with a frequency $\Omega > 1$. Engels {\it et al} created  a giant hole in a rapidly rotating dilute BEC by removing atoms from the rotating condensate using a tightly focused resonant laser \cite{Engels}.  We, on the other hand, show that a giant hole can be created  in the condensate by a cluster of impurities. For this we place a cluster of impurities, a central impurity surrounded by five other impurities, at the center of the system. The effective potential (trap + impurity) is shown in Fig.  \ref{Fig. 4}(a). It may be noted that in contrast to the case in \cite{Kasamatsu2} where the trap potential is a `Mexican hat', in our case the effective potential is a `Mexican hat' together with five peaks along the circular ring at the bottom of the hat.  The condensate density in this effective potential in absence of rotation is shown in Fig. \ref{Fig. 4}(b). Fig. \ref{Fig. 5}(a) shows a giant hole at the center of the rotating condensate. On comparison of Fig. \ref{Fig. 5}(a) with Fig. \ref{Fig. 2}(a) (the undistorted vortex lattice) we see that in Fig. \ref{Fig. 5}(a) three visible vortices are missing. However the average angular momentum of these two vortex lattices remains almost same, namely, $<L_z>=21.4$ for lattice in Fig. \ref{Fig. 5}(a) and $<L_z>=20.8$ for lattice in Fig. \ref{Fig. 2}(a). It may be mentioned here that according to Feynman rule $<L_z> \sim N/2$ where $N$ is the number of vortices in the system. If there are both visible and hidden vortices in the condensate then $<L_z> \sim (N_v+N_h)/2$ where $N_v$ and $N_h$ denote the total number of visible and hidden vortices respectively in the system \cite{dey14}. On careful examination we observed that there are three hidden vortices inside the giant hole. The presence of three hidden vortices at the center of the giant hole can be seen from the phase profile of the condensate density as shown in Fig. \ref{Fig. 5}(b).  Fig. \ref{Fig. 5}(b) shows that there are $N_v=34$ visible vortices and $N_h=3$ hidden vortices and therefore the total number of vortices ($N_v+N_h=37$) in  Fig. \ref{Fig. 5}(a) is same as the number of visible vortices $N_v=37$ in Fig. \ref{Fig. 2}(a). Fig. \ref{Fig. 5}(c) shows that a giant hole can be created at off center positions with similar phase singularities. This is similar to the experimentally observed  off center giant vortices \cite{Engels}. It may be noted that in contrast to \cite{Kasamatsu2} in our case the giant hole can be created even for $\Omega < 1$ (Fig. \ref{Fig. 5}(d)). Our calculations show  that the size of the giant hole increases with increasing strength of the impurity potential and also with  increasing size of the impurity cluster.

In order to see the effect of random impurities on the vortex lattice we generate random potential at each spatial grid points uniformly distributed over $[-V_0,V_0]$, where $V_0$ is the strength of the random potential \cite{Ghosal}. Fig. \ref{Fig. 6}(a-d) shows the vortex lattice configuration for various 
strength of the random potential for a particular realization of the disorder.
From the figures we can see that the vortex lattice structure gets increasingly disordered with increasing strength of the impurity potential eventually leading to the melting of the lattice structure (Figs. \ref{Fig. 6}(d)). The vortex lattice  in Fig. \ref{Fig. 6}(d) has reached almost random disorder. To show that the long-range order or translational periodicity is destroyed we calculate the structure factor and the histogram of the distances between each pair of vortices of the melted vortex lattice in Fig. \ref{Fig. 6}(d). The results are shown in Fig. \ref{Fig. 7}.  The diffused structure factor profile in Fig. \ref{Fig. 7}(a) shows that there is no regular structure of the vortex lattice in Fig. \ref{Fig. 6}(d). Figs. \ref{Fig. 7}(b) and \ref{Fig. 7}(c) shows the histogram plot for the undistorted vortex lattice (Fig. \ref{Fig. 2}(a)) and the melted vortex lattice (Fig. \ref{Fig. 6}(d)) respectively. The visibility of well separate peaks in Fig. \ref{Fig. 7}(b) reveals the high degree of long-range order of the corresponding undistorted vortex lattice in Fig. \ref{Fig. 2}(a). Similarly, the bare visibility of the well separated peaks of the histogram plot in Fig. \ref{Fig. 7}(c) shows that there is no long range-order in the melted vortex lattice in Fig. \ref{Fig. 6}(d). Fig. \ref{Fig. 8}(a) shows the melted vortex lattice for $\Omega < 1$. Similarly, the melted vortex lattice for the harmonic trap potential is shown in \ref{Fig. 8}(b). We now show that the vortex lattice can also be melted by disorder or the pseudorandom potential created by the superposition of two optical lattices as mentioned above. Fig. \ref{Fig. 9}(a) shows the melted vortex lattice in presence of the pseudorandom bichromatic optical lattice potential $V_{imp} = V_0[\cos(2\pi\beta x + \phi) + \cos(2\pi\beta y + \phi)] $ and Fig. \ref{Fig. 9}(c) shows its corresponding histogram plot. Similarly, Fig. \ref{Fig. 9}(b) shows the melted vortex lattice in presence of the pseudorandom potential $V_{imp} = V_0[\cos^2(kx) + \cos^2(ky)] + V_1[ \cos^2(\textbf{k}_1 . \textbf{r})+\cos^2(\textbf{k}_2 . \textbf{r})]$ obtained by adding an additional optical lattice to the main optical lattice as mentioned above and  Fig. \ref{Fig. 9}(d) shows its corresponding histogram plot. Again, the absence of non-separated peaks in the histogram plots show the melting of the vortex lattice  in presence of the pseudorandom potential created by the superposition of optical lattices. If the disorder is removed, then the system goes back to the Abrikosov triangular lattice with long range order or tanslational periodicity. 

In conclusion, we have demonstrated the disorder induced vortex lattice melting in a rotating Bose-Einstein condensate. The origin of the melting of vortex lattice is  the pinning of vortices by the impurity as demonstrated above. We have shown that even a single impurity can distort the vortex lattice and the vortex lattice gets pinned to the impurity for sufficient strength of the impurity potential. The distortion of the vortex lattice also depends on whether the impurity position is commensurate or incommensurate with the positions of the vortices in the undistorted (Abrikosov) vortex lattice. The vortex lattice distortion is maximum when the impurity is placed in the center of the two vortices of the Abrikosov vortex lattice. There is however a small distortion of the vortex lattice even if the impurity is placed exactly on the Abrikosov vortex lattice position (commensurate position). This is due to the mismatch between the sizes of the impurity and the Abrikosov vortices. We further show that a new type of giant hole with hidden vortices inside can be created in the vortex lattice by a cluster of impurities. In presence of random impurities or disorder the vortices gets  pinned at random positions leading to melting of the vortex lattice.  Similarly, in presence of pseudorandom potential or disorder created by the superposition of two optical lattices, the vortices get pinned to the random optical lattice sites resulting in the melting of the vortex lattice.  Since the random potential in BEC can be realized with available experimental techniques, we feel that the disorder induced vortex lattice melting and other results presented in the manuscript can be observed experimentally. 

B. D thanks DST and BCUD-SPPU for the financial support through  research projects. K.P. thanks the NBHM, IFCPAR, DST-FCT and CSIR, Government of India, for the financial support through major projects. BD would like to thank M. Karmakar for useful discussions.

\end{document}